# High Thermoelectric Performance of Au@Sb$_2$Te$_3$ Heterostructure Derived from the Potential Barriers

Wenwen Zheng [a], Peng Bi [b], Fengming Liu [c], Yong Liu [b], Jing Shi [b], Rui Xiong [*, b], Ziyu Wang [*, d, e]

[a] School of Science, Wuhan Institute of Technology, Wuhan 430205, China
[b] Key Laboratory of Artificial Micro-and Nano-structures of Ministry of Education and School of Physics and Technology, Wuhan University, Wuhan, 430072, China
[c] School of Mathematics and Physics, China University of Geosciences, Wuhan 430074, China
[d] College of Physics and Electronic Science, Hubei Normal University, Huangshi 435002, China
[e] Dongfeng Commercial Vehicle Technology Center, Wuhan, 430056, China
* Address correspondence to zywang@whu.edu.cn and xiongrui@whu.edu.cn
†Electronic supplementary information (ESI) available.

## Abstract

The correlated couple of electrical and thermal property is the challenge to realize a substantial leap in thermoelectric materials. Synthesis of semiconductor and metal composites is a significant and versatile design strategy to optimize the thermoelectric performance driven by tailored interface between nanoinclusions and matrix. In this study, we present the simultaneous increase of electrical conductivity and Seebeck coefficient, and reduction of thermal conductivity in Sb$_2$Te$_3$ − Au system. The enhanced electrical conductivity lies in the incorporated Au nanostructures contributing to injecting carriers to Sb$_2$Te$_3$ matrix. The appropriate barriers originated from the Au − Sb$_2$Te$_3$ interface, which filter low energy carriers, results in enhancement of Seebeck coefficient. The increased boundaries and nanodomains block the transport of phonons, subsequently reducing the thermal conductivity. As a consequence, combination of these effects promote double of ZT value in 1% Au@ Sb$_2$Te$_3$ composites with respect to the pristine Sb$_2$Te$_3$.

## 1. Introduction

Thermoelectric (TE) materials can realize the mutual conversion between electric energy and thermal energy based on Seebeck effect and Peltier effect.[1] To make TE materials into practical application, the efficiency is determined by the dimensionless figure of merit *ZT*, defined as $ZT = \frac{S^2 \sigma}{\kappa} T$, where $S, \sigma, \kappa, T$ are the Seebeck coefficient, electrical conductivity, thermal conductivity and absolute temperature, respectively. So an ideal TE material with large Seebeck coefficient, electrical conductivity and low thermal conductivity is the target that researchers persist to pursue. However, ZT still remain around 1 owing to the trade-off between the three TE parameters. Since Hicks and Dresselhaus predicted the low-dimensional materials, such as quantum wire and quantum well, exhibited low thermal conductivity and enhanced TE performance,[2, 3] researchers' interests have been rekindled. Nanostructuring and low dimensional approach has been proven to be an effective strategy to improve ZT via enhancing boundary scattering of phonons to dramatically decrease the lattice thermal conductivity.[4-8] Recently, Biswas *et al.* proposed a panoscopic approach to fabricate all-scale hierarchical architectures by taking advantages of endotaxial nanostructuring, mesoscale grain boundaries and atomic scale substitutional doping in p-type PbTe. The ZT value of ~2.2 at 915 K revealed the role of the nanostructuring in controlling phonon transport of bulk thermoelectric materials.[9]

Solution synthesis is a scalable bottom-up chemical route to design nanostructured grains with controlled composition, sizes and morphologies.[10-15] Various nanostructures, such as nanoplatelets, nanowires, and nanotubes have been fabricated through this low-cost and facile process. Through the application of chemical synthesis, Lei Yang *et al.* fabricated $Cu_2Se$ nanoplates,[16] Yuho Min obtained $Bi_2Se_3$ and $Bi_2Te_3$ nanoplates,[17] Qun Wang designed PbTe nanowire with branched nanorods,[18] Sook Hyun Kim prepared $Bi_2Te_3$ nanotubes by the interdiffusion of Bi and Te metals.[19] Though such nanostructured materials exhibiting low thermal conductivity owing to scattering phonons from numerous grain boundaries and interfaces, the electrical conductivity is also deteriorated arising from carriers being simultaneously scattered and suppressed. Thus, the solution derived bulk materials usually show a low ZT value compared to that by ball milling and hot pressing approaches due to the low power factor. Ajay Soni *et al.* reported the chemically synthesized $Bi_2Te_3$ nanoplatelets, exhibiting the maximum ZT of 0.1 at 300 K.[14] How to enhance the electrical conductivity of the solvothermal method fabricated TE materials with augment performance is a tough issue to be addressed. Developing multi-phase composite is an emerging effective strategy to improve the property in solution processed TE materials.[20] Theories have proven that the second phase, when integrated at proper length scale, can be shorter than the phonon mean free path but longer than that of electron, thus resulting in simultaneously scattering the heat carrying phonons and favoring the transport of the carriers. However, when choosing semiconductor or oxides as the embedded NPs into the TE matrix, the electrical property is negatively affected owing to the carriers scattering by the incorporated NPs.[21] Li-Dong Zhao *et al.* studied the effects of nanostructured SiC nanoparticles (NPs) on the thermoelectric performance of n type $Bi_2Te_3$, the NPs dispersed composites exhibiting increased Seebeck coefficient and decreased electrical conductivity.[22] Incorporation of metal nanoparticles with controlled sizes into the semiconductor matrix, forming a metal-semiconductor heterostructure, is an effective way to optimize the electrical property and increase ZT values.[23, 24] When the size and distribution ratio is suitable tuned, the introduced metal can simultaneously increase the electrical conductivity through optimizing the carrier concentration and enhance the Seebeck coefficient based on the low energy electron filtering effect in which electrons with low energy are selectively filtered by the bending band between the interface of nanoparticle.[25, 26] Qihao Zhang *et al.* fabricated a hierarchical two-phased heterostructure by exotically introducing silver nanoparticles into $Bi_2Te_3$ matrix, resulting in an improvement of power factor due to the excellent electrical transport property of Ag and enhanced Seebeck coefficient.[27] However, silver is apt to be oxidized when exposed in air. In comparison gold is an alternative metal beyond silver to hybrid into TE matrix to obtain high TE performance and good stability. The gold nanoparticles (Au NPs) with defined nanostructures have excellent electrical conductivity, stability, high aspect ratio and large surface areas, which is widely applied in areas of biosensing, phononics and electrics, etc.[28-30]

P type antimony telluride ($Sb_2Te_3$) based bulk TE materials dominate the low-temperature power and cooling application. Though nanostructured $Sb_2Te_3$ are synthesized via solution method with controlled morphology and structures, the organic agents are hard to remove from the surface of nanoparticles, leading to the poor electrical conductivity and ZT value below 1. Constructing hierarchically Au @$Sb_2Te_3$ heterostructured bulk composites sets forth a new avenue to promote the optimization of TE performance. Au has a low working function (~ 5.31 eV),[31] which enables injection of carriers into $Sb_2Te_3$ semiconductor. Adjusting the fraction of Au nanostructures, the interface of metal – semiconductor is tuned to align the potential barrier, resulting in a remarkable increase of electrical conductivity. Nanostructured Au domains and increased boundaries helps to boost phonons scattering, thus decreasing the thermal conductivity. The simultaneous combination of enhanced electrical conductivity, large Seebeck coefficient, and reduced thermal conductivity is highly beneficial to optimizing ZT value of Au@$Sb_2Te_3$ composites.

Herein we report the Au@$Sb_2Te_3$ heterostructures via in situ reducing Au NPs on solvothermal synthesized hexagon nanoplatelets. The mean diameter of Au NPs is ~ 10 nm at the concentration of 1%, while the nanoparticles grow up comparable to 20 nm at concentration higher than 3%. It is essentially stressed that the high conductive Au NPs is decorated uniformly in the $Sb_2Te_3$ matrix, realizing in a synchronously regulating the electrical and thermal property. The conductive Au NPs induced electron injection, energy filtering effect and nanoinclusion-matrix boundary scattering, as a whole, which contribute to increasing the electrical conductivity, enhancing the Seebeck coefficient and reducing the thermal conductivity at the meantime. Consequently, a maximum figure of merit of ZT reaches 0.8 for the sample containing 1% Au NPs, which is 180% enhancement compared to 0.38 for the pristine $Sb_2Te_3$.

## 2. Materials and methods

2.1. Synthesis method

In a typical synthesis process, $Sb_2Te_3$ nanoplatelets were fabricated via solvothermal method with a stoichiometric of $K_2TeO_3$, $SbCl_3$, NaOH and polyvinylphrrolidone (PVP). After dissolving the raw materials in the solution of diethylene glycol (DEG) and reacting at 240 °C for 3 h at a 250 ML round bottom flask, the dark suspensions of $Sb_2Te_3$ were transferred and washed with acetone and isopropyl alcohol several times. The as prepared $HAuCl_4$/$AgNO_3$/citrate mixture solution was injected into the $Sb_2Te_3$ matrix in the boiling water for 15 minutes with stirring to ensure homogenous reaction. The reacted solution was further refluxed under stirring for 2 h to guarantee the formation of uniform Au NPs on the $Sb_2Te_3$ nanoplateletes.

2.2. Characterization

X-ray diffraction (XRD) patterns were carried out by using a Bruker D8 advance diffractometer with Cu Ka radiation. The microstructures characterization of Au@$Sb_2Te_3$ heterostructure was performed on scanning electron microscopy (SEM, HITACHI S-4800), atomic force microscopy (AFM) as well as transmission electron microscope (TEM, JEM-2010). The Raman spectra of the as prepared samples were taken on Confocal Raman Micro spectroscopy (RM-1000, Renishaw) with 514 nm excitation laser wavelength. The temperature dependence of electrical conductivity and Seebeck coefficient were measured on a commercial equipment (ZEM-3, ULVAC-RIKO) in a He atmosphere from 300 K to 523 K. The thermal conductivity $\kappa$ was calculated with the relationship $\kappa=D\rho C_p$, where $D$ is thermal diffusivity, $\rho$ is density and $C_P$ is specific heat. The thermal diffusivity $D$ was tested on Netzsch LFA 457 by laser flash diffusivity method. The density $\rho$ was measured by Archimede method and $C_p$ was determined by differential scanning calorimetry apparatus (DSC-Q50). The Hall mobility $\mu$ and effective carrier concentration $n$ were calculated from the relationship $\mu=\sigma/ne$ and $n=1/R_He$.

## 3. Results and discussion

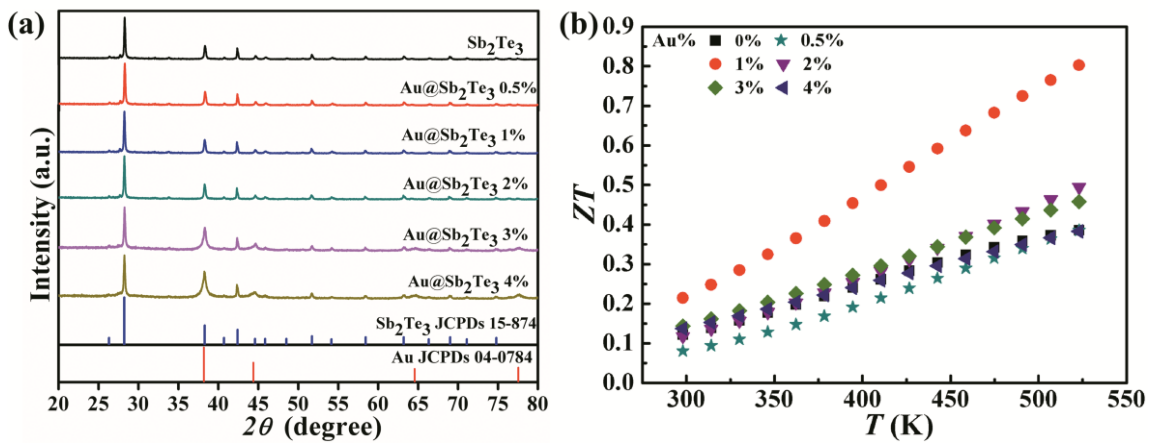

**Figure 1.** (a) Powers XRD patterns of $Sb_2Te_3$ at different concentrations of Au NPs (0, 1%, 2%, 3%, 4%). (b) Temperature dependence of ZT value for pure $Sb_2Te_3$ and Au@$Sb_2Te_3$ composites.

The phase and crystal structure of pristine $Sb_2Te_3$ and the Au@$Sb_2Te_3$ composites at different concentrations are characterized by the XRD patterns and the spectrum are shown in Fig. 1a. The diffraction peaks of the database $Sb_2Te_3$ (JCPDs #15-875) and Au (JCPDF #04-0784) are provided in vertical line as references. The peaks of pure sample $Sb_2Te_3$ match well with the standard card, and the diffraction peaks of the Au@$Sb_2Te_3$ composites with low addition of Au (<3%) show little change with respect to $Sb_2Te_3$ due to the intensity below the instrument detection. There is an overlap between the second strongest peak of $Sb_2Te_3$ and the highest peak of Au. The main diffraction peak of Au becomes obvious when the concentration of Au excesses 3%, demonstrating the existence of Au. The components of the Au-$Sb_2Te_3$ heterostructures are further confirmed by energy-dispersive X-ray spectroscopy (EDS) in Fig. S1.

The temperature dependence of ZT value of pure $Sb_2Te_3$ and Au@$Sb_2Te_3$ between 300 K and 523 K is shown in Fig. 1b. Compared with Au-free $Sb_2Te_3$, ZT value of the heterostructure is enhanced from 0.38 to 0.45 for 3% Au, and to the maximum of 0.8 with the optimized Au concentration of 1%, which is twice of the pristine. It is addressed that the incorporated Au nanoparticles decorate the microstructure of $Sb_2Te_3$ to form metal-semiconductor heterostructure and engineer the transport path of carriers and phonons. Tuning the ratio of Au to an optimal value allow the optimization of carrier concentration and construction of energy filtering barrier, thus decouple the electrical conductivity and Seebeck coefficient, resulting in a remarkable increase of power factor. Furthermore, the incorporated interface and boundaries in the multiphase strongly scatter phonons to suppress the thermal conductivity. The detailed information are discussed below. As a result, the trade-off between the three parameters ($\sigma$, S, $\kappa$) is broken owing to the Au-$Sb_2Te_3$ assembly heterostructure, realizing double of ZT value at 523 K.

The temperature dependence of the electrical and thermal transport properties of the samples are shown in Fig. 2. As presented in Fig. 2a, the electrical conductivity of the Au-free $Sb_2Te_3$ and Au@$Sb_2Te_3$ composites decrease with the temperature, indicating a semimetal behavior. The electrical conductivity behaves a rising trend with addition of Au concentration, increasing from $2.04 \times 10^4$ S/m to $3.89 \times 10^4$ S/m at 523 K as the ratio of Au up to 4%, except for a fluctuating value of $1.75 \times 10^4$ S/m at 0.5% Au NPs. This is maybe due to the incorporation of minor NPs acting as the carrier block, hindering the transport of electrons. When the concentration of Au reaches a percolation (1%), the Au NPs provide extra electrons to multi-phase system, which compensate for the carrier scattering effect. As a result, the electrical conductivity performs higher value when the content of Au is above 0.5%.

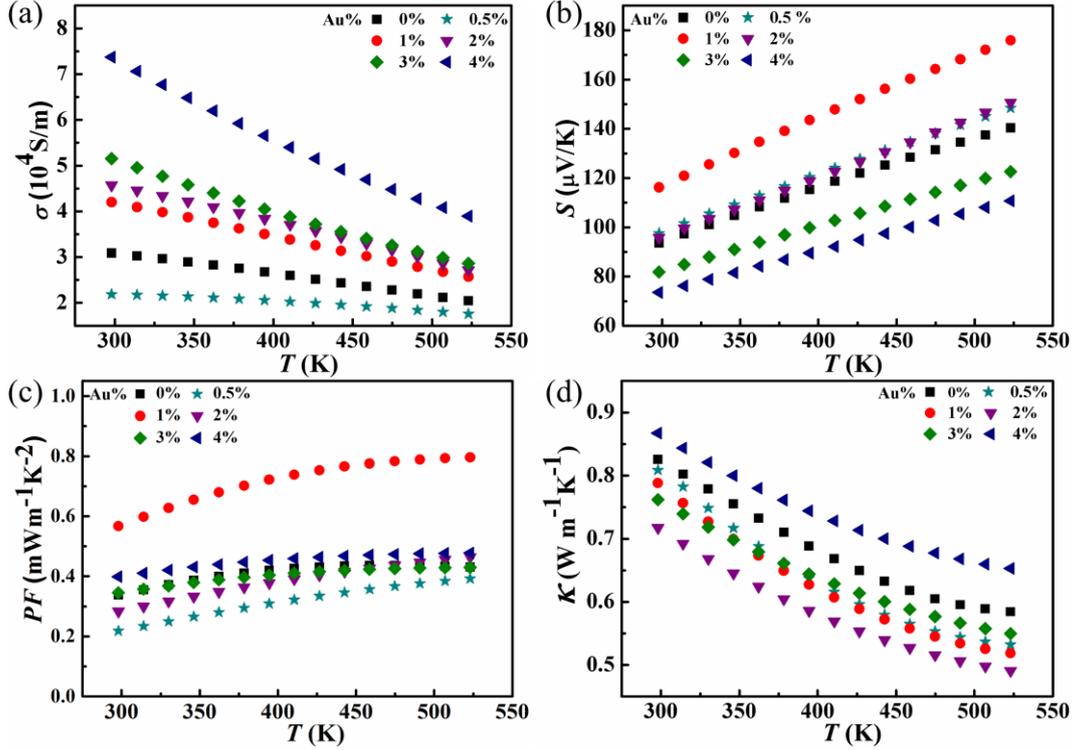

**Figure 2.** Temperature dependence of electrical conductivity (a), Seebeck coefficient (b), power factor (c), and thermal conductivity (d) of the pure $Sb_2Te_3$ and $Au@Sb_2Te_3$ composites.

The temperature dependence of Seebeck coefficient is displayed in Fig. 2b. Both pure phase $Sb_2Te_3$ and $Au@Sb_2Te_3$ composites have positive Seebeck coefficient, revealing that the majority carriers are holes and the samples are p-type, which is in accordance with the Hall measurement data (Fig. S2). Compared to the pristine $Sb_2Te_3$, the Seebeck coefficient of the composites increase with the Au content, reaching the maximum of 175.8 μV/m at 1% ratio, which is enhanced by 25%. The artificially formed Au-$Sb_2Te_3$ heterostructure develops a schottky barrier, which selectively filters carriers with low energy, however, permits the high energy carriers to pass through, generating an energy filtering effect which is responsible for the increased Seebeck coefficient. The Seebeck coefficient starts to decrease from 150.6 μV/K to 110.7 μV/K when the ratio of Au is above 1%, For degenerate semiconductors (parabolic band, energy independent scattering approximation), the Seebeck coefficient is given by $S = \frac{8\pi^2 \kappa_B^2}{3eh^2} m^* T \left(\frac{\pi}{3n}\right)^{2/3}$, where $m^*$ is the effective mass of the carriers and $n$ is the carrier concentration. This relationship reflects that the Seebeck coefficient is inversely proportional to the carrier concentration. The addition of Au dramatically increases the carrier concentration of the composites, thus leading to a reduction of Seebeck coefficient. This result demonstrates that the Seebeck coefficient can be tuned to an optimized value upon the optimal combination of Au and $Sb_2Te_3$, which is dependent on the ratio and distribution of Au NPs.

According to the measured electrical conductivity and Seebeck coefficient, the power factor is calculated as shown in Fig. 2c. Owing to the simultaneous enhancement of electrical conductivity and Seebeck coefficient with addition of 1% Au NPs, the power factor is improved significantly up to 0.79567 mW m$^{-1}$K$^{-2}$ in comparison to 0.43078 mWm$^{-1}$K$^{-2}$ of Au-free sample, increased by 84.7%.

The total thermal conductivity for different samples as functions of temperature are shown in Fig. 2d. The thermal conductivity of all the samples display the same trend, monotonously decreasing with the temperature. In contrast with the pure $Sb_2Te_3$, the addition of 0.5% Au NPs lead a reduction of thermal conductivity from 0.58439 Wm$^{-1}$K$^{-1}$ to 0.53225 Wm$^{-1}$K$^{-1}$ at 523 K. This value is further decreased to 0.51849 Wm$^{-1}$K$^{-1}$ for the sample containing 1% Au, afterwards reaches the minimum of 0.49083 Wm$^{-1}$K$^{-1}$ at 2% Au content. However, we find that the excess addition of Au results in an increase of thermal conductivity of 0.54973 Wm$^{-1}$K$^{-1}$ for the sample with 3% Au, and finally achieving 0.65288 Wm$^{-1}$K$^{-1}$ when the concentration of Au reaches 4%.

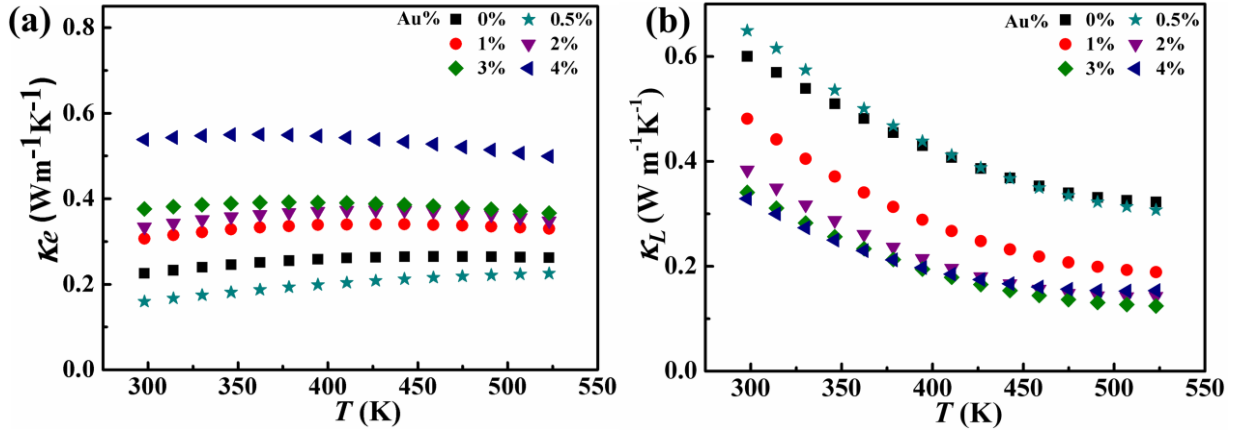

**Figure 3.** Temperature dependence of electrionic thermal conductivity (a) and lattice thermal conductivity (b).

To evaluate the electronic and lattice contribution to the total thermal conductivity, temperature dependence of electronic thermal conductivity $\kappa_e$ and lattice thermal conductivity $\kappa_L$ is present in Fig. 3. The electronic thermal conductivity is calculated based on the Wiedmann-Franz law: $\kappa_e = L_0 T \sigma$, where $\kappa_e$ is the electronic thermal conductivity, $L_0$ is the Lorenz number, $T$ is the absolute temperature, $\sigma$ is the electrical conductivity. The lattice thermal conductivity is obtained by subtracting the electronic component from $\kappa$. It can be seen that the electronic thermal conductivity of the composites firstly decrease at low Au concentration 0.5%, then monotonously increase with adding Au and reach the maximum of 0.499 Wm$^{-1}$K$^{-1}$ for the sample with 4% Au. This trend is in accordance with that of the electrical conductivity owing to the correlationship of $\sigma$ and $\kappa_e$. The excessive increase of conductive Au makes a significant enhancement of electrical conductivity, thus leads to an increase of $\kappa_e$. The lattice thermal conductivity of the composite at 0.5% Au concentration is comparable to Au-free sample, and decrease with the ratio of Au. This is mainly due to the introduction of Au NPs acting as the phonon scattering centers, rendering the phonons with short and medium mean free paths scattered. Meanwhile, the strong Au-Sb$_2$Te$_3$ grain boundary scattering also makes contributions to the reduction of $\kappa_e$. As a consequence, the lattice thermal conductivity is dramatically decreased, reaching the lowest value 0.12 Wm$^{-1}$K$^{-1}$ at 523 K, which is reduced by 61% with respect to the pure phase. In order to further explore the effect of boundary and interface on reducing thermal conductivity in Au@Sb$_2$Te$_3$ composites, we make a calculation of thermal conductivity by using the lattice thermal conductivity of the pristine Sb$_2$Te$_3$, which ignore the role of interface scattering. A comparison of thermal conductivity is made between the experimental and calculated results (shown in Fig. S3). The results indicate that experimental thermal conductivity is far below the calculation results, owing to interface and grain boundary scattering.

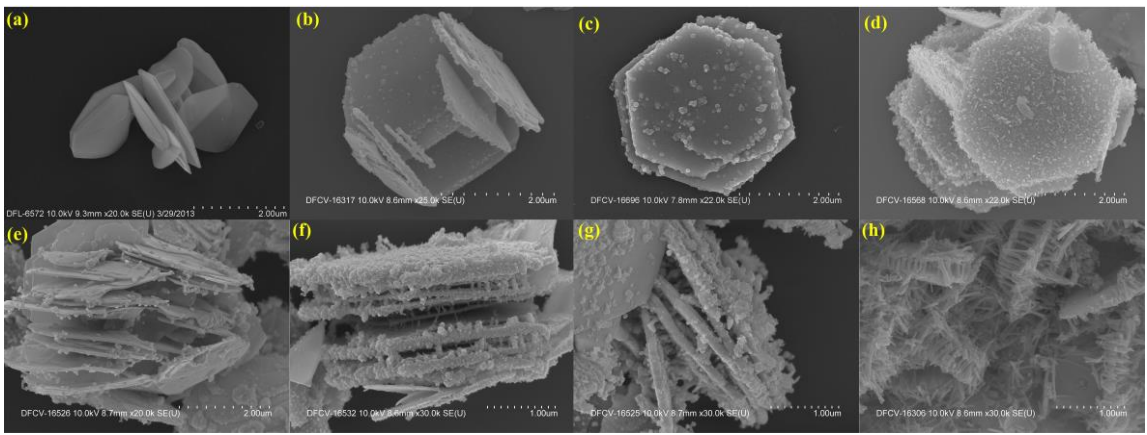

**Figure 4.** FESEM micrograph of pure Sb$_2$Te$_3$ (a), and Au@Sb$_2$Te$_3$ composites at different mole ratio of Au 1% (b), 3% (c) and 4% (d). The Au nanoparticles decorate on the Sb$_2$Te$_3$ platelets and interface (d)-(f).

Fig. 4 shows the representative microstructure of the as-prepared Sb$_2$Te$_3$ pure phase and the Au@Sb$_2$Te$_3$ composites. As shown in Fig. 4a, the pristine Sb$_2$Te$_3$ nanoparticles synthesized via solvothermal method are observed to be hexagonal nanosheets with the edge about 1~2 μm and thickness several nm. Au nanoparticles are grown on the nanoplatelets of Sb$_2$Te$_3$ by HAuCl$_4$-assisted citrate reduction. The obtained Sb$_2$Te$_3$ nanosheets with flat surfaces and regular edges are acted as the heterostructuring nucleation and seeding positions for Au nanoparticles. The formed Au nanoparticles are several nm spheres uniformly dispersed on Sb$_2$Te$_3$. Few Au nanoparticles are detected to form heterostructures at low mole concentration (1%). With the increase of Au concentration, the relative distribution of gold nanoparticles becomes dense and agglomerated. A higher concentration (3%) leads to more decorated Au dots with larger size. As the concentration enhanced to be 4%, the whole Sb$_2$Te$_3$ nanoplatelets are thickly coated by Au nanoparticles on the surfaces and edges. Some Au particles connect with each other, thus resulting in the electrical and thermal properties of the composites strongly influenced by this heterostructure.

Fig. 4d-f shows the decoration of Au on the surface and between the interlayers of Sb$_2$Te$_3$. Some gold NPs distributed uniformly on the Sb$_2$Te$_3$ platelet to coat a thin conductive layer as shown in Fig. 4d, while others are inserted around the edges of layers to bridge conducting channels as shown in Fig. 4e-f. This heterostructure plays a significant role on the electrical property and Seebeck coefficient. The selecting Au phase can allow effectively injection of charge carriers into the Sb$_2$Te$_3$ and the formation of the Au@Sb$_2$Te$_3$ heterostrucure facilitate the carrier tunneling. It should be mentioned that the coherent interfaces of Au and Sb$_2$Te$_3$ can strongly scatter the low energy electrons, resulting in the low energy filtering effect, which prescribes for enhancement of Seebeck coefficient.

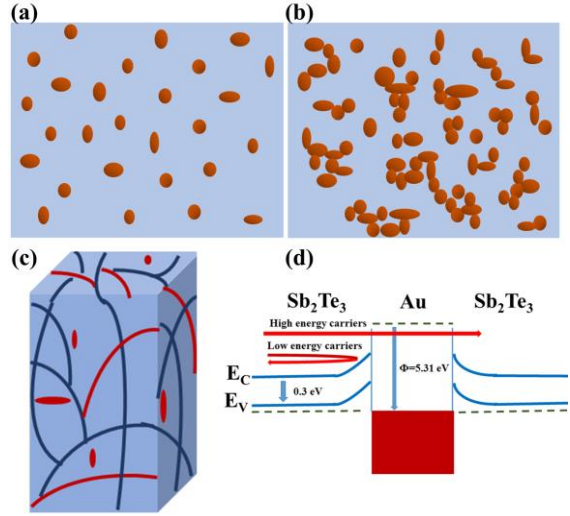

**Figure 5.** Schematic of (a) a small amount of Au decoration on the Sb$_2$Te$_3$ nanoplatelets, (b) plenty of Au nanoparticles distribution on the matrix, (c) densified Au@Sb$_2$Te$_3$ composites with boundaries, (d) band alignment which results in the barrier formed between the interface of Au nanostructures and Sb$_2$Te$_3$ semiconductor.

The distribution of Au nanoparticles are depicted in Fig. 5a and 5b. A small amount of Au nanodots are scattered on the Sb$_2$Te$_3$ host, which are nanocrystal building blocks to hinder the transport of electrons and phonons, thus decreasing the electrical conductivity and thermal conductivity. When the Au nanoparticles increase to the threshold value, some Au nanodomains are connected together as electron path, which inject carriers to the Sb$_2$Te$_3$ matrix. This is why the Au@Sb$_2$Te$_3$ composites present a monotonous increase trend in electrical conductivity when the Au content rises above 1%. Figure 5c shows the compacted Au@Sb$_2$Te$_3$ pellets fabricated by spark plasma sintering. The doped Au nanoinclusion results in the bend of Sb$_2$Te$_3$ band structure (Fig. 5d). Sb$_2$Te$_3$ is a degenerate semiconductor, whose Fermi level is exactly located inside the valence band. The working function $\phi_{Au}$ of Au is 5.31 eV, while the ionization potential $\phi_{Sb_2Te_3}$ of Sb$_2$Te$_3$ is 4.45 eV [32]. The interface between Sb$_2$Te$_3$ host and Au nanocrystals forms an energy barrier $V_B$ ($V_B = \phi_{Au} - \phi_{Sb_2Te_3}$), which is approximately 0.86 eV, making the low energy carriers block back and let the high energy carriers pass through. This energy filtering effect increase the average energy of carriers, resulting in the increase of Seebeck coefficient of composites when the Au content is below 1%.

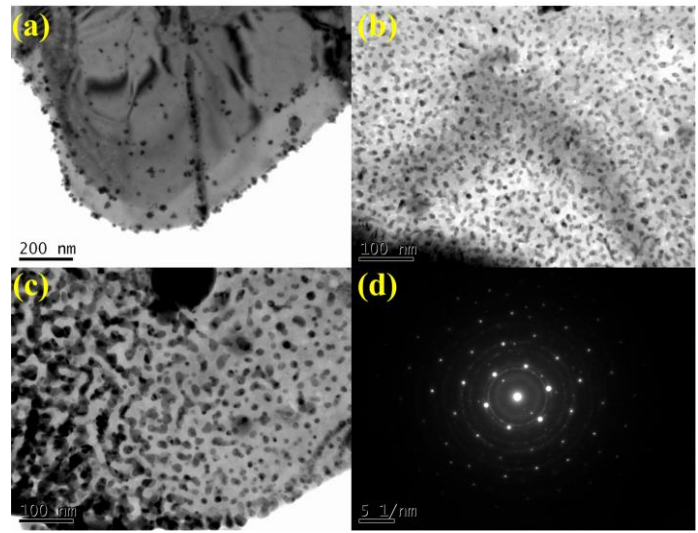

**Figure 6.** TEM images of the as-prepared Au@Sb$_2$Te$_3$ heterostructure (a-c). Au NPs with size of ~10 nm are uniformly coated on the surface of Sb$_2$Te$_3$. (d) SAED pattern reveals the existence of the two phase of Au and Sb$_2$Te$_3$.

Transmission electron microscopy (TEM) image further reveals the presence of spherical or ellipsoidal Au nano particles decorated on the Sb$_2$Te$_3$ matrix, forming a metal-semiconductor heterostructure with coherent interface, which is shown in Fig. 6. The selected area electron diffraction (SAED) patterns are indicative of Sb$_2$Te$_3$ with rhombohedral (space group R$\bar{3}$m) lattice structure and Au with face-centered cubic structure. The Au NPs are widespread and distributed on the host, with diameters of ~10 nm, which make contributions to the transport of electrons and phonons. The diameter of the Au NPs is larger than electrons with short and medium mean free paths, but smaller than phonons of the medium and long mean free paths. Around 80% of the lattice thermal conductivity is attributed from phonon modes with medium and long mean free paths [33]. This heterostructure can effectively scatter heat-carrying phonons, thereby leading to a low thermal conductivity. More interestingly, the carriers are not significantly affected. At the optimal $Sb_2Te_3$/Au ratio, the electrical conductivity is increased to some extent. This is due to the compensation of the electron injection from Au NPs.

How the thermoelectric performance of the composites increase with the concentration of Au at 523 K is present in Fig. 7. The simulation of electrical conductivity, Seebeck coefficient, thermal conductivity and ZT value is applied using the classical percolation power law as shown in Fig. 7 a-d. The percolation threshold is estimated with this law [34, 35]: $\psi = \psi_0 \left| (f - f_{Au})/f \right|^{-q}$, where $\psi$ is the thermoelectric parameter of Au@$Sb_2Te_3$ heterostructure, $\psi_0$ is the thermoelectric parameter of $Sb_2Te_3$ matrix, $f$ is the molar ratio of Au at the threshold, $f_{Au}$ represents the molar ratio of Au@$Sb_2Te_3$ composite and $q$ is the power law exponent. The best fit of the electrical conductivity to the percolation theory gives the electrical percolation threshold of 0.67%, which indicates the turning of electrical conductivity for the composites upon 0.67% molar ratio of Au. This result is in consistent with the experimental data, in which the electrical conductivity have a slight decrease at 0.5%, afterwards increase with the concentration of Au. In the same way, the fitted threshold of Seebeck coefficient is 1.2%, predicting the optimal Seebeck coefficient of the composite reach at 1.2% Au. This trend is obviously observed in Fig. 7b with the experimental data and simulation coincident. The addition of Au with moderate ratio induces an energy filtering barrier between $Sb_2Te_3$ and Au which scatters the carriers with low energy and benefits to the enhancement of Seebeck coefficient. We also take the percolation simulation on the thermal conductivity, which has a threshold value of 2.1%. The incorporated Au NPs can simultaneously scatter the carriers and phonons, thus resulting a decrease of thermal conductivity. However, this would be offset by the increase of $\kappa_e$ with the addition of Au due to the increase of electrical conductivity. The turning point of thermal conductivity depends on the concentration of Au. Significantly, the percolation theory gives a well assess of the optimal concentration of Au. As a consequence, a percolation threshold of 1.2% is obtained for ZT value as shown in Fig. 7d. The fitted results demonstrate the optimization of thermoelectric performance is obtained at low ratio of Au, which agrees well with the experimental data.

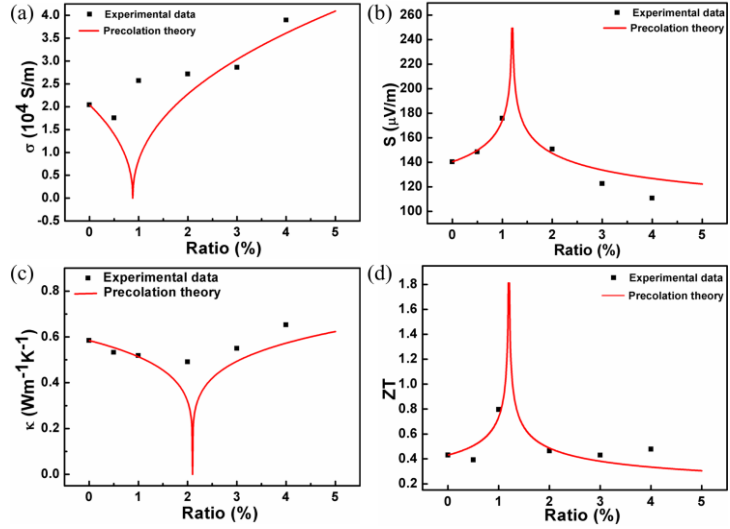

**Figure 7.** Comparison of experimental and theoretical thermoelectric performance using percolation theory for (a) electrical conductivity, (b) Seebeck coefficient, (c) thermal conductivity, (d) ZT value.

## 4. Conclusion

In summary, Au@Sb$_2$Te$_3$ heterostructures have been synthesized in a scalable and controlled solution method. The Au NPs exhibited a distribution of 10 nm, which uniformly dispersed on the Sb$_2$Te$_3$ nanoplatelets to form Au-Sb$_2$Te$_3$ heterostructures. The ratio of the mixing Au NPs is tuned to achieve simultaneous enhancement of electrical conductivity and Seebeck coefficient. Meanwhile, the thermal conductivity is effectively decreased owing to the scattering phonons by the incorporated Au NPs. As a consequence, ZT of the composite with 1% Au reaches the optimal value 0.8 at 523 K. Percolation theory predicted that thermoelectric performance of Au@Sb$_2$Te$_3$ heterostructures can be optimized by the concentration of incorporated Au. This provides an optional manufacturing technique to design multiphase TE materials with high conductive metal for excellent thermoelectric materials.

## Acknowledgements

This work was supported by National Natural Science Foundation of China (Grant No. 11504103, No. 11474224, No.11474225, No. 51571152, No. 11674088, No. 11704289).